\documentclass[aps,pra,onecolumn,superscriptaddress,longbibliography,showkeys]{revtex4-2}
\usepackage{multirow}
\usepackage{wrapfig}
\usepackage{geometry}
\usepackage{ragged2e}
\usepackage{enumitem}
\usepackage{amsmath}
\usepackage{amssymb}
\usepackage{graphicx}
\usepackage{amsthm}
\usepackage{amsfonts}
\usepackage{color}
\usepackage{epstopdf}
\usepackage{epsfig}
\usepackage{mathrsfs}  
\usepackage{bussproofs}	 
\usepackage{caption}
\usepackage{orcidlink}

\newtheorem{Xeorem}{Theorem}[section]
\newtheorem{definition}[Xeorem]{Definition}
\newtheorem{remark}[Xeorem]{Remark}
\newtheorem{oml-poset}[Xeorem]{\bf Historical Quantum Conventions.}
\newtheorem{dist-conv}[Xeorem]{\bf Historical Classical Conventions.}
\newtheorem{lemma}[Xeorem]{Lemma}

\newtheorem{claim}[Xeorem]{Claim}
\newtheorem{response}[Xeorem]{Response}
\usepackage{float}

\usepackage{xcolor}

\DeclareRobustCommand{\crossover}[2]{%
  \mathrel{\ooalign{%
    \hfil$#1$\hfil\cr
    \hfil$#2$\hfil\cr
  }}%
}

\def\crta{\vrule height1.41ex depth-1.27ex width0.34em}
\def\dj{d\kern-0.36em\crta}
\def\Crta{\vrule height1ex depth-0.86ex width0.4em}
\def\Dj{D\kern-0.73em\Crta\kern0.33em}
\begin{document}


\title[Logics with Multiple Nonisomorphic Models]{Classical and Quantum Logics with Multiple Non-Equivalent and Non-Isomorphic Models}[]

\author{Mladen Pavi\v ci\'c \orcidlink{0000-0003-1915-6702}\ }
\email{mpavicic@irb.hr}
\homepage{https://mladen-pavicic.github.io/web/}
\affiliation{Institute Ru{\dj}er Bo\v skovi\'c, 10000 Zagreb, Croatia}
\affiliation{Institute of Physics, 10000 Zagreb, Croatia}

\begin{abstract}
  Both quantum and classical logics have several mutually
  non-equivalent and non-isomorphic models as has repeatedly been
  proved in the literature. Yet, several authors have recently
  come with criticisms of this result. In this paper, we show that
  the criticisms are either unfounded or wrong. 
\end{abstract}

\maketitle

\section{\label{Sec.~Intro}Introduction}

In \cite{mpcommp99} we show that there are non-isomorphic and
non-equivalent models for quantum as well as for classical logic.
In \cite{pavicic-amp-16} we enlarge
the number of such models. Models represent the semantics of
logic and in textbook presentations they are divided in several
interpretations. For instance, in \cite[Ch.~8-10]{Schechter},
they are grouped in numerically valued, set-valued, and
topological interpretations. The set-valued interpretations
include powerset, our O6 (called hexagon)
\cite[p.~272, 305]{Schechter}, crystal, and Church's diamond ones.

Yet, some authors and their referees are of the opinion that 
certain characterizations of our interpretations of quantum and
classical logics are inappropriate, in particular Kowalski, Paoli
and Giuntini \cite{kowalski-paoli-giuntini-16} and Drozdowska
\cite{drozdowska-26}. Therefore, in this paper, we address
the discrepancies between their characterizations of the
considered interpretations and the characterizations accepted by
others in the literature and by ourselves. In doing so,
we have to start with notions, concepts, definitions, systems,
theorems, procedures, and results we already put forward in
several of our previous papers but that were misinterpreted
by the aforementioned critics and possibly other readers.
That being our path, we can only do our best to present those
facts in a condensed juxtaposed manner, quantum vs.~classical,
aiming to pinpoint that there is nothing contradictory in
having non-equivalent and non-isomorphic models of a single
considered logic. Thereupon we explicitly address the criticisms
in Sec.~\ref{Sec.~crit}.

The paper is organized as follows.

In Sec.~\ref{Sec.~QandCL} quantum and classical
logic are presented in a comparative manner.

In Sec.~\ref{Sec.~QandCLa} various weakly orthomodular,
orthomodular, weakly distributive, and distributive lattices
are presented in a similar way.

In Sec.~\ref{Sec.~models} we present mutually non-equivalent and
non-isomorphic models of quantum and classical logical and proofs
of soundness and completeness for all of them.

In Sec.~\ref{Sec.~crit} we present two criticisms of the fact that
quantum and classical logics have mutually non-equivalent and
non-isomorphic models recently put forward by several authors. We
analyse each of their claims and show that they are either
unfounded or provably wrong.

In Sec.~\ref{Sec.~concl} we summarize our arguments against the
criticisms. 

\section{\label{Sec.~QandCL}Quantum and Classical Logics}

Logic ($\mathcal{L}$) is a language  defined by a set of
axioms and rules imposed on propositions. We consider quantum
$\mathcal{QL}$ and classical $\mathcal{CL}$ axiomatic logics.

Primitive (elementary) propositions/well-formed formulae (wffs)
in $\mathcal{L}$ are denoted as $p_0,p_1,p_2,...$; primitive
connectives are $\neg$ (negation) and $\vee$ (disjunction).
$p_j$ is a wff for $j=0,1,2,...$; $\neg A$ is a wff if $A$ is a wff;
$A\vee B$ is a wff if $A$ and $B$ are wffs. We denote the set of all
wffs by ${\mathcal{F}}^\circ$. Further operations are
defined as follows.

\begin{definition}\label{D:conj}Conjunction:
$\qquad A\wedge B\ {\buildrel\rm def\over =}\ \neg (\neg A\vee\neg B).$
\end{definition}

\begin{definition} \label{def:impl-0} Classical implication:
$\qquad A\to_{\rm c} B\ {\buildrel\rm def\over =}\ \neg A\vee B.$
\end{definition}

\begin{definition} \label{def:impl-3} Quantum implication:
$\quad A\to_q B\ \ {\buildrel\rm def\over =} \ \
(\neg A\wedge \neg B)\vee(\neg A\wedge B)\vee(A\wedge(\neg A\vee B)).\quad$
\end{definition}

\begin{definition} \label{def:impl-1} Sasaki implication:
$\qquad A\to_1 B\ {\buildrel\rm def\over =}\ \neg A\vee (A\wedge B).$
\end{definition}

\begin{definition}\label{L:equiv} Quantum equivalence:
$\qquad A\equiv_{q} B\ \
{\buildrel\rm def\over =}\ \ (\neg A\wedge\neg B)\vee(A\wedge B).$
\end{definition}

\begin{definition}\label{L:equiv-0} Classical equivalence:
$\qquad A\equiv_{\rm c} B\ \
{\buildrel\rm def\over =}\ \ (B\to_{\rm c} A)\wedge(A\to_{\rm c} B).$
\end{definition}

\begin{definition}\label{L:derive}
By ${\mathcal{F}}=\langle {\mathcal{F}}^\circ,\neg,\vee\rangle$
we denote logical system built within $\mathcal{L}$ by means of a
set of all wffs $\mathcal{F}^\circ$ containing $\vee$ and $\neg$
via axioms and rules of inference so as to yield expressions
called theorems (axioms are also theorems). The set of theorems
is denoted by $\vdash$;  $A\in\ \vdash$ means that $A$ is a
theorem. We write it as $\vdash A$ to mean ``$A$ is provable,''
i.e., if $A$ is a theorem, then there is a proof of it.
\end{definition}

\begin{definition}\label{L:derive+}
For $\Gamma\subsetneq{\mathcal{F}}^\circ$ we say $A$ is derivable
from $\Gamma$ and write $\Gamma\vdash_{\mathcal{L}} A$ or just
$\Gamma\vdash A$ if there is a finite sequence of
formulae, the last of which is $A$, and
each of which is either one of the axioms of
$\mathcal{L}$ or is a member of\/ $\Gamma$ or is obtained from its
precursors with the help of a rule of inference of the logic.
\end{definition}

We do not give all the axioms here. The reader can find them in
\cite{pavicic-amp-16}.  Notice that quantum logic requires a lot
more axioms than classical logic but that is not of importance
for the current presentation.

\noindent
\begin{minipage}[t]{0.48\textwidth}
\setlength{\jot}{2pt}
\begin{align}
&{\boldsymbol{\rm Quantum\ Logic: \mathcal{QL}}}\nonumber\\
&Axioms:\nonumber\\
&\vdash A \vee B \equiv_q B \vee A\nonumber\\
&\vdash A \equiv_q \neg\neg A\nonumber\\
&\dots\nonumber\\
&Rule\ of\ Inference:\nonumber\\
&\vdash A\ \&\ \vdash A \rightarrow_q B
\Rightarrow \vdash B\nonumber
\end{align}
\end{minipage}%
\hfill
\begin{minipage}[t]{0.48\textwidth}
\setlength{\jot}{2pt}
\begin{align}
&{\boldsymbol{\rm Classical\ Logic: \mathcal{CL}}}\nonumber\\
&Axioms:\nonumber\\
&\vdash A \vee B \to_c B \vee A\nonumber\\
&\vdash A \to_c \neg\neg A\nonumber\\
&\dots\nonumber\\
&Rule\ of\ Inference:\nonumber\\
&\vdash A\ \&\ \vdash A\ \to_c B
\Rightarrow \vdash B\nonumber
\end{align}
\end{minipage}

\section{\label{Sec.~QandCLa}Quantum and Classical
  Lattices}

A {\em lattice\/} $L=\langle L^\circ,\cup,\cap\rangle$
is a partially ordered set such that for every pair of
$a,b\in L^\circ$ there exist both  $a\cap b\in L^\circ$ and
$a\cup b\in L^\circ$. Properties of intersection and union in the
algebra of sets are carried over into $L$ as expressions called
conditions of $L$. The partial order of elements $a\le b$
acquires the forms $a\cup b=b$ and $a\cap b=a$. When the negation
``$\>'\>$'' is added we obtain an {\em ortholattice\/}
$OL=\langle OL^\circ,',\cup,\cap\rangle$ and when the
weak-orthomodularity-like condition {\it{w-oml-like-c}} is added to
an OL we obtain a weakly orthomodular lattice WOML (in general a
class of WOMLs). Notice that the greatest ($a\cup a'$) and the least
($a\cap a'$) elements of a lattice, usually denoted as 1 and 0,
respectively, need not be numbers.

Again we do not give all the lattice conditions here. The reader
can find them in \cite{pavicic-amp-16}.

        \begin{align}
{\boldsymbol{\rm Weakly}}&\ {\boldsymbol{\rm Orthomodular\ Lattice\ WOML:}}\nonumber\\
 &Conditions:\nonumber\\
     &a \cup b = b \cup a\nonumber\\
&a = a''\nonumber\\
     &\dots\nonumber\\
          {\text{\it w-oml-like-c}}\quad &
a\cup(b\cap(a'\cup b'))\equiv_{q}a\cup b=1
   \label{eq:bareWOML}
        \end{align}

By adding the following conditions we obtain quantum and
classical lattices. OML stands for orthomodular lattice,
WDOL for weakly distributive ortholattice, and WD for
distributive lattice (Boolean algebra). In general for
classes of such lattices.

\raggedbottom

\noindent
\begin{minipage}[t]{0.48\textwidth}
\vspace{0pt}
\setlength{\jot}{2pt}
\begin{align}
&\boldsymbol{\rm Quantum\ Lattices\!:}
\nonumber\\
\nonumber\\
&\boldsymbol{\rm WOML_1}=\boldsymbol{\rm WOML}\ +
\nonumber\\
&[(a\equiv b)'\to_1a']\ =\ (a\to_1b)
\nonumber\\
\nonumber\\
&\boldsymbol{\rm OML}=\boldsymbol{\rm WOML}\ +
\nonumber\\
&a\cup(b\cap(a'\cup b'))=a\cup b
\nonumber\\
&(orthomodularity)
\nonumber\\
\nonumber\\
&{\rm OML\subsetneq WOML_1\subsetneq WOML}
\nonumber
\end{align}
\end{minipage}%
\hfill
\begin{minipage}[t]{0.48\textwidth}
\vspace{0pt}
\setlength{\jot}{2pt}
\begin{align}
&\boldsymbol{\rm Classical\ Lattices\!:}
\nonumber\\
\nonumber\\
&\boldsymbol{\rm WDOL}=\boldsymbol{\rm WOML}\ +
\nonumber\\
&a\cup(b\cap c)\equiv_c(a\cup b)\cap(a\cup c)=1.
\nonumber\\
\nonumber\\
&\boldsymbol{\rm DL\ (Boolean\ Alg.)}=\boldsymbol{\rm WOML}\ +
\nonumber\\
&a\cup(b\cap c)=(a\cup b)\cap(a\cup c)
\nonumber\\
&(distributivity)
\nonumber\\
\nonumber\\
&{\rm DL\subsetneq WDOL\subsetneq WOML}
\label{woml-incl}
\end{align}
\end{minipage}

\addtocounter{equation}{1}
        
\medskip
        
A sublattice of $L$ which does not contain any stronger lattice or
in our case any of of the considered stronger lattices we denote by
$L^*$.

\begin{align}
{\rm DL,\ OML, \ WDOL\ \not\subset WOML^*}\label{no-woml}\\
{\rm DL,\ OML \not\subset WOML_1^*}\label{no-woml1}\\
{\rm DL \not\subset WDOL^*}\label{no-wdol}
\end{align}

In \cite{pavicic-amp-16} we provide some examples of these
lattices (e.g., O6 and O7). 

\section{\label{Sec.~models}Lattices as Models of
  Logics}

\begin{definition}\label{exists-c}We call ${\mathcal{M}}=\langle
L,h\rangle$ a model of of a set of wffs $\Gamma$ from logic
$\cal L$ if $L$ is a lattice and a valuation
$h:{\mathcal{F}}^\circ\longrightarrow L$  is a morphism of
formulae ${\mathcal{F}}^\circ$ from $\cal L$ into $L$ which
satisfy $h(A)=1$ for any $A\in\Gamma$ and preserve
the operations $\neg,\vee$ while turning them into $',\cup$;
we say that $A$ is true in the model $\mathcal{M}$.
\end{definition}

\begin{definition}\label{one}We call a formula $A\in{\mathcal{F}}^\circ$
valid in the model $\mathcal{M}$, and write $\vDash_{\mathcal{M}} A$, if
$h(A)=\textstyle{1}$ for all valuations $h$ on the model, i.e.,
for all $h$ associated with the base set $L$ of the model. 
We call a formula $A\in{\mathcal{F}}^\circ$ a consequence of\/ 
$\Gamma\subsetneq{\mathcal{F}}^\circ$ in the 
model $\mathcal{M}$ and write $\Gamma\vDash_{\mathcal{M}} A$ 
if $h(X)=\textstyle{1}$ for all $X$ in\/ $\Gamma$ implies 
$h(A)=\textstyle{1}$, for all valuations $h$.
\end{definition}

\subsection[Subsec.~soundness]{Soundness}

To prove soundness means to prove that the axioms and rules of
inference and consequently all theorems of $\mathcal{L}$ hold
in the models of $\mathcal{L}$. The models of $\mathcal{QL}$
are O6, WOML*, WOMLi*,  and OML, and of $\mathcal{CL}$
are O6, WOML, WDL* and DL. 

With the help of valuation $h$, given by Def.~\ref{exists-c}, 
which maps elements of logic to elements of lattice $L$: 
$a=h(A)$, $b=h(B)$, \dots and operations $\neg,\vee$
to $\neg,\vee$, we can prove all mapped axioms and 
the rule of inference in any model $\mathcal{M}=L$,
where $L$ can be WOML, WOML*, OML, WDOL, WDOL*, DL, etc. 

\begin{Xeorem}\label{th:soundness}{\rm [Soundness]}
  $\qquad\Gamma\vdash_{\mathcal{L}}
  A\quad\Rightarrow\quad\Gamma\vDash_{\mathcal{M}}A$
\end{Xeorem}

\subsection[Subsec.~completeness]{Completeness}

Completeness means that every lattice formula which has a
valuation ``1'' that holds in all of the lattice models is
provable in the logic.

\begin{Xeorem}\label{th:completeness}{\rm[Completeness]}
$\qquad\Gamma\vDash_\mathcal{M} A\qquad\Rightarrow\qquad
\Gamma\vdash_{\mathcal{L}} A$.
\end{Xeorem}

For each of the models there is a corresponding instance of
completeness. For example, for the weakly orthomodular
lattice not containing the axiom of orthomodularity WOML*:
$\Gamma\vDash_\mathcal{\rm WOML*}
A\Rightarrow\Gamma\vdash_{\mathcal{QL}} A$, for the orthomodular
lattice model OML: $\Gamma\vDash_\mathcal{\rm OML}
A\Rightarrow\Gamma\vdash_{\mathcal{QL}} A$, for the weakly 
distributive lattice not containing the axiom of distributivity
model WDOL*: $\Gamma\vDash_\mathcal{\rm WDOL^*}
A\Rightarrow\Gamma\vdash_{\mathcal{CL}} A$,  for the distributive
lattice model DL: $\Gamma\vDash_\mathcal{\rm DL}
A\Rightarrow\Gamma\vdash_{\mathcal{CL}} A$, etc.
Notice that two non-equivalent models WOML* and OML
(non-orthomodular and orthomodular) generate the same quantum
logic $\mathcal{QL}$ and that two non-equivalent models WDOL* and
DL (non-distributive and distributive) generate the same classical
logic $\mathcal{CL}$.

Technically, we carry the completeness proofs with the help of
the Lindenbaum algebras. In order to carry them out we have to
introduce some definitions and lemmas.

\begin{definition}\label{D:equiv-class-sets-woml}
The equivalence class for a wff $A$ under the relation of 
equivalence $\approx$ given by 
Eq.\ (\ref{eq:equiv-noml-re}) is defined as $|A|=\{B\in 
{\mathcal{F}}^\circ:A\approx B\}$, and we denote
${\mathcal{F}}^\circ/\!\approx\ =\{|A|:A\in {\mathcal{F}}^\circ\}$.
The equivalence classes define the natural morphism
$f:{\mathcal{F}}^\circ\longrightarrow
{\mathcal{F}}^\circ/\!\approx$, which gives
$f(A)\ =^{\rm def}\ |A|$. We write $a=f(A)$, $b=f(B)$, etc.
\end{definition}

However, the relation $a=b$ on ${\mathcal{F}}^\circ$ which
defines a relation of equivalence $\approx$
\begin{eqnarray}
  A\approx B \qquad\Leftrightarrow\qquad |A|=|B|
  \qquad\Leftrightarrow\qquad \Gamma\vdash A\equiv B
\label{eq:equation-non-om-q}
\end{eqnarray}
would encompass all formulas WOML contains, including those
from OML. Hence, we have to change the definition of the
relation of equivalence so as to make it impossible for both 
$a\cup(b\cap(a'\cup b'))$ and $a\cup b$ to be in the same
class of equivalence since that would amount to
$a\cup(b\cap(a'\cup b'))=a\cup b$, i.e., to the
orthomodularity. We achieve that by expanding the definition
of the relation of equivalence by a term which verifies
that the equation can pass a chosen WOML lattice.  
We have chosen lattice O6 in which the orthomodularity fails
and all WOML equations and rules pass. 

\begin{figure}[ht]
\center
\includegraphics[width=0.98\textwidth]{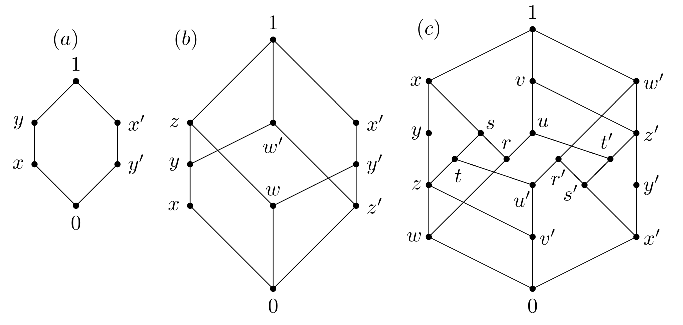}
\caption{(a) O6; (b) O7; (c) O8.}
\label{fig:Os}
\end{figure}

\begin{definition}\label{D:hexagon}Letting O6
represent the lattice from Figure \ref{fig:Os}, we define 
${\mathcal{O}}${\rm 6} as the set of all mappings 
$o6:{\mathcal{F}}^\circ\longrightarrow {\rm O6}$ 
such that for $A,B\in{\mathcal{F}}^\circ$,
$o6(\neg A)=o6(A)'$, and 
$o6(A\vee B)=o6(A)\cup o6(B)$.
\end{definition}

\begin{Xeorem}\label{th:congruence-nonoml}
The relation of {\em equivalence} $\approx_{\Gamma,\mathcal{QL}}$
or just $\approx$, defined as
\begin{eqnarray}\label{eq:equiv-noml-re} 
A\approx B\ \ {\buildrel\rm
def\over =}\ \
\Gamma\vdash
A\equiv B\ (\forall o6\in
{\mathcal{O}6})\{[(\forall X\in\Gamma)(o6(X)=
\textstyle{1})]\Rightarrow [o6(A)=o6(B)]\}\hbox to 30pt{}
\end{eqnarray}
is a relation of congruence, where\/ 
$\Gamma\subsetneq{\mathcal{F}}^\circ$
\end{Xeorem}

Since $o6(A\vee(B\wedge(\neg A\vee \neg B)))=x$ and
$o6(A\vee B)=y$, meaning that the orthomodularity violates O6,
the relation of equivalence, so defined, excludes the OML and
confirms the WOML$^*$ as a model of quantum logic. 

\begin{lemma}\label{L:lind-alg-non-q} Lindenbaum algebra
${\mathcal{A}}=\langle {\mathcal{F}}^\circ/\!\approx,\neg/\!\approx,
\vee/\!\approx\rangle=\langle {\mathcal{F}}^\circ/\!\approx~,f\rangle$
is a {\rm WOML*} model of $\Gamma$ in which the {\rm WOML*}
conditions hold for $\neg/\!\approx$ and $\vee/\!\approx$ as  $'$
and $\cup$ respectively.
\end{lemma}

\begin{Xeorem}\label{th:completeness-wom}(Completeness)
If a formula {\rm A} is true in all {\rm WOML} models of a set of
wffs $\Gamma$, i.e., if $f(A)=1$, then $\Gamma\vdash A$.
\end{Xeorem}

\begin{proof}$f(A)=1$ is equivalent to $|A|=|B\vee\neg B|$ and
therefore to
\begin{align}
&&\Gamma\vdash A\equiv_i B\vee\neg B\quad\&\quad (\forall o\in{\cal O}6)
[(\forall X\in\Gamma)(o(X)=1)\ \Rightarrow\ o(A)=1]\nonumber\\
&&\Leftrightarrow\Gamma\vdash A\quad\&\quad (\forall o\in{\cal O}6)
[(\forall X\in\Gamma)(o(X)=1)\ \Rightarrow\ o(A)=1]\nonumber\\
&&\Rightarrow\Gamma\vdash A\label{last}\nonumber\hbox to 226pt{}\qedhere
\end{align}
\end{proof}
\begin{Xeorem}\label{th:compl-sound-wom}
  $\Gamma\vdash_{\mathcal QL} A$ iff $A$ is true in all {\em WOML*}
  models.
\end{Xeorem}

There are WOML conditions that are stronger than those given
by Eq.~(\ref{eq:bareWOML}), e.g., the one that defines WOML1
above or the following one which defines WOML2.
\begin{eqnarray}\label{eq:woml2}
  [(a\to_1b)\equiv(b\to_1a)]=(a\equiv b)
\end{eqnarray}

Here we should give Lemma 22 from \cite{pavicic-amp-16} together
with its proof because the proof from
\cite[Lemma 22]{pavicic-amp-16} contains several typos.

\begin{lemma}\label{lemma:woml0-2}
 {\rm OML} is properly included in (i.e., it is stronger than) 
{\rm WOML2}, {\rm WOML2} is properly included in {\rm WOML1}, 
and {\rm WOML1} is properly included in {\rm WOML}.
{\rm WOML*} is a WOML which contains neither OML, nor WOML2, nor
WOML1 conditions nor any other condition stronger than WOML's ones.
{\rm WOML1*} is a WOML which contains neither OML nor WOML2
conditions nor any other condition stronger than WOML1's one but
contains the WOML1 condition.
{\rm WOML2*} is a WOML which contains OML condition nor any other
condition stronger than WOML2's one but contains the WOML1 and
WOML2 conditions.
\end{lemma}

\begin{proof} The orthomodularity condition fails
in O6, O7, and O8 from Fig.~\ref{fig:Os}. The condition which
defines WOML2 (Eq.~(\ref{eq:woml2})) passes O6, but fails in
O7 and O8. The condition which defines WOML1 passes O6 and O7,
but fails in O8. 
\end{proof}

We also obtain
\begin{Xeorem}\label{th:compl-sound-womm}
  $\Gamma\vdash_{\mathcal QL} A$ iff $A$ is true in all
  {\em WOML1,2*} models.
\end{Xeorem}
{\parindent=0pt as well as}
\begin{Xeorem}\label{th:compl-sound-womm-bare}
  $\Gamma\vdash_{\mathcal QL} A$ iff $A$ is true in all
  {\em WOML,1,2} models. {\rm [Without ``*'']}
\end{Xeorem}

For classical logic we have \cite{pavicic-amp-16}

\begin{Xeorem}\label{th:compl-sound-womcl}
  $\Gamma\vdash_{\mathcal CL} A$ iff $A$ is true in all
  {\em WOML} models.
\end{Xeorem}

\begin{Xeorem}\label{th:compl-sound-wdol}
  $\Gamma\vdash_{\mathcal CL} A$ iff $A$ is true in all
  {\em WDOL*} models.
\end{Xeorem}

{\parindent=0pt as well as}

\begin{Xeorem}\label{th:compl-sound-womcl-bare}
  $\Gamma\vdash_{\mathcal CL} A$ iff $A$ is true in all
  {\em WOML, WDOL} models.
\end{Xeorem}

This is all in addition to the standard models: 

\begin{Xeorem}\label{th:compl-sound-oml}
  $\Gamma\vdash_{\mathcal QL} A$ iff $A$ is true in all
  {\em OML} models.
\end{Xeorem}

\begin{Xeorem}\label{th:compl-sound-Boole}
  $\Gamma\vdash_{\mathcal CL} A$ iff $A$ is true in all
  {\em DL (Boolean algebra)} models.
\end{Xeorem}

\begin{figure}[ht]
\center
\includegraphics[width=0.98\textwidth]{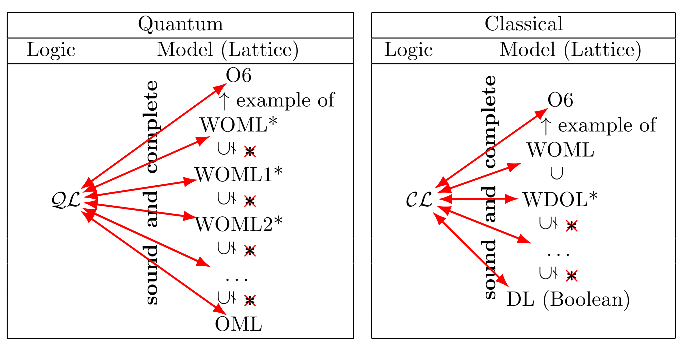}
\captionsetup{justification=justified, singlelinecheck=false}
\caption{Both quantum and classical logics have multiple
  mutually non-isomorphic and non-equivalent models. Sign
  $\crossover{\textcolor{red}{\boldsymbol{\times}}}{\boldsymbol{*}}$
  means that, e.g., WOML* does not contain WOML1*
  (WOML1* $\not\subset$ WOML*) while WOML does contain WOML1
  (WOML1 $\subsetneq$ WOML). Cf.~Eqs.~(\ref{no-woml}), (\ref{no-woml1}),
  and (\ref{no-wdol}) vs.~(\ref{woml-incl}).}
\label{fig:two-tables}
\end{figure}

The theorems are given a graphical representation in
Fig.~\ref{fig:two-tables}.

\section{\label{Sec.~crit}Remarks, Criticisms, and
     Responses}

In our papers  \cite{mpcommp99,pavicic-amp-16} we made the
following remark:

\begin{remark}\label{remark1}
  In quantum logic $\mathcal{QL}$ there is nothing orthomodular
  and in classical logic $\mathcal{CL}$ there is nothing
  distributive. 
\end{remark}

What we meant by this we explained as follows (ibid.):
\begin{remark}\label{remark2} [Ibid.]
The following theorems hold in $\mathcal{QL}$ and $\mathcal{CL}$:
\begin{eqnarray}
&\vdash_{\mathcal{QL}} A\vee(B\wedge(\neg A\vee \neg B))
\equiv_q  A\vee B,\label{oml-equiv}\\
&\vdash_{\mathcal{CL}}A\vee(B\wedge C)\equiv_c (A\vee B)
  \wedge (A\cup C).\label{d-equiv}
\end{eqnarray}
Their lattice mapping in  WOML are:
\begin{eqnarray}
&a\cup(b\cap(a'\cup b'))\equiv_q a\cup b=1,\label{omw}\\
&a\cup(b\cap c)\equiv_c (a\cup b) \cap (a\cup c)=1.\label{dw}
\end{eqnarray}
However in OML and DL, because of 
\begin{eqnarray}
  {\rm OML}\ axiom\!: \qquad &a\equiv_q b=1\quad \Rightarrow\quad
    a=b,\label{om=1}\\
  {\rm DL}\ axiom\!: \qquad &a\equiv_c b=1\quad  \Rightarrow\quad
    a=b,\label{d=1}
\end{eqnarray}
  we additionally have 
\begin{eqnarray}
&a\cup(b\cap(a'\cup b'))=a\cup b,\label{om}\\
&a\cup(b\cap c)=(a\cup b) \cap (a\cup c),\label{d}
\end{eqnarray}
which are traditionally called {\em the orthomodularity} and
{\em the distributivity}.
\end{remark}

Therefore, what we meant by Remark \ref{remark1} is explained by
\ref{remark2}. In the literature, the orthomodularity
and distributivity are usually expressed in the equational form
like in Eqs.~(\ref{om=1}, \ref{d=1}) and it is our opinion that
the equivalence forms given by Eqs.~({\ref{oml-equiv},
\ref{d-equiv}) should not be called the orthomodularity and 
distributivity, respectively. This is substantiated by the
following conventions which, in effect, reduce quantum logic
to an orthomodular lattice and classical logic to the Boolean
algebra while we formulate them as genuine logics. 

\begin{oml-poset}\phantom{.}\\
  ``Quantum logic is a poset which possesses a least and a greatest
  element.'' {\rm\cite[p.~1]{ptak-pulm} [A poset with unique least
  upper bound (join, $a \cup b$) and a unique greatest lower bound
  (meet, $a \cap b$) is a lattice.]}\\
  ``An orthomodular poset is a quantum logic\dots Sometimes in the
  literature \dots quantum logic is understood {\rm [to be]} a
  lattice.''  {\rm \cite[p.~73]{dvurecenskij-book}}.
\end{oml-poset}

\begin{dist-conv}\phantom{.}\\
  ``{\rm [In classical logic]} there is \dots\ a distributive law
  $X\wedge (Y\vee Z)\ {\rm eq.}\ (X\wedge Y)\vee(X\wedge Z)$
  {\rm [where]}  {\rm ``eq.''}  does not belong to {\rm [the]} set
  of logical symbols.''
  {\rm \cite[pp.~6, 7]{hilb-ack-book}}\\
  ``{\rm [The distributivity]} is expressed as {\rm [given by
  Eq.~(\ref{dw})];} $\equiv_c$ is defined by the truth table.''
  {\rm \cite[pp.~29, 32]{copi}}  
\end{dist-conv}

We see that directly or indirectly the authors in the field use
the terms defined in the models of a logic, not in the logic
itself. This seems to have misled Kowalski, Paoli and Giuntini
\cite{kowalski-paoli-giuntini-16} into the following claims. 

\begin{claim}[Kowalski, Paoli and Giuntini]{\it
  ``It has been taken for granted for a long time that orthomodular
  lattices are the “algebraic counterpart” of orthomodular quantum
  logic. Pavi\v ci\'c and Megill have questioned this claim by
  pointing out that orthomodular quantum logic is sound and complete
  with respect to a proper supervariety of the variety {\rm OML} of
  orthomodular lattices (the so-called weakly orthomodular
  lattices {\rm [WOML]}).''
  {\rm\cite[Abstract]{kowalski-paoli-giuntini-16}}}
\label{KPG1}
\end{claim}

\begin{response} {\it to Claim \ref{KPG1}}. Notice that the authors
  call the quantum logic (``the oldest and noblest,'' p.186)
  orthomodular quantum logic ({\em idem per idem} argument in an
  attempt to prove that quantum logic itself is orthomodular). 
  They further express their criticism as follows. 
\end{response}

\begin{claim}[Kowalski, Paoli and Giuntini]{\it 
\dots lesson to be drawn from Pavi\-\v ci\'c's and Megill’s
proof. {\rm WOML} falls short of the requirements needed for
an equivalent algebraic semantics for $\mathcal QL$  precisely
because it includes algebras (like {\rm O6}) containing as members
distinct but $\mathcal QL$-indiscernible elements. And this happens
because WOML does not universally satisfy {\rm [Eq.~(\ref{om=1})]}
which characterises orthomodular lattices within the class of all
ortholattices (as Pavi\v ci\'c and Megill correctly observe).
Therefore,  Pavi\v ci\'c’s and Megill’s semantics is not a good
semantics in the sense we specified above.
{\rm\cite[p.~189]{kowalski-paoli-giuntini-16}}}
\label{KPG2}
\end{claim}

\begin{response} {\it to Claim \ref{KPG2}}.
Apparently, the essence of the completeness proof of all
$\mathcal QL$ models escapes the authors. To repeat, completeness
means that every lattice formula which has a valuation “1”
that holds in all of the lattice models is provable in the
logic, meaning that all models WOML, WOML*, WOML1,\dots OML yield
the same theorems of $\mathcal QL$ (``indiscernible elements'')
although these models are mutually non-isomorphic and
non-equivalent as we show in the previous sections above and
graphically present in Fig.~\ref{fig:two-tables}.
It seem that the authors are not at ease with this result.
\end{response}

\begin{claim}[Kowalski, Paoli and Giuntini] {\it\
We disregard whatever differences there may be between logical
and algebraic types.
{\rm\cite[p.~182]{kowalski-paoli-giuntini-16}}}
\label{KPG3}
\end{claim}

\begin{response} {\it to Claim \ref{KPG3}}.
As a result, it seems that the authors make use of the same
variable letters (e.g., $\alpha$ and $\beta$) for elements of
logic and of its models (lattices).
\end{response}

\begin{claim}[Kowalski, Paoli and Giuntini] {\it\
Lemma 3.1. A logic $\mathcal L$ is algebraisable with equivalent
algebraic semantics $K$ iff it satisfies AL1:
$\Gamma\vdash_{\mathcal L}\alpha\ {\text{iff}}\
\tau(\Gamma)\vdash_{Eq(K)}\tau(\alpha)$
and AL4: $\alpha\approx\beta \dashv\vdash_{Eq({\mathcal{L}})}
\tau(\rho(\alpha,\beta))$.\ 
{\rm\cite[p.~185]{kowalski-paoli-giuntini-16}}}
\label{KPG4}
\end{claim}

\begin{response} {\it to Claim \ref{KPG4}}.
The meaning of AL1 and AL4 is (more or less) clear from the
next claim. The definitions of the symbols are to be looked for
in their paper. 
\end{response}

\begin{claim}[Kowalski, Paoli and Giuntini] {\it\
$\mathcal QL$ is strongly sound and complete w.r.t.~OML. This
settles \dots AL1 \dots [And] AL4 [means that we have] 
$\alpha\approx\beta \dashv\models\alpha\equiv\beta\approx 1$.
However, it is known that $\alpha\equiv\beta\approx
1\Rightarrow\alpha\approx\beta$ holds in OML.
{\rm \cite[p.~182]{kowalski-paoli-giuntini-16}}
\dots the unique position of OML as the algebraic match of
$\mathcal QL$ is secure\dots its algebraic counterpart can’t be
anything else than the variety of orthomodular lattices.
Surprisingly enough, however, this claim has been questioned by
Pavi\v ci\'c and Megill in a series of papers. 
{\rm\cite[p.~188]{kowalski-paoli-giuntini-16}}}
\label{KPG5}
\end{claim}

\begin{response} {\it to Claim \ref{KPG5}}.
Recalling Claim \ref{KPG3}, here it seems that the authors assume
that $\alpha$ and $\beta$ are elements of the lattice and that
$\approx$ means $=$. As for ``questioning,'' we have rigorously
proved that $\mathcal QL$ has not only OML as its model/semantics
but also many other non-equivalent and non-isomorphic WOML models.  
\end{response}

\begin{claim}[Kowalski, Paoli and Giuntini] {\it\
\dots it is \dots comparatively easy to detect the flaws in the
argument by Pavi\v ci\'c and Megill. What they manage to show,
indeed, is that deductive quantum logic (as they call it)
{\rm [$\mathcal CL$]} has the variety of weakly orthomodular
lattices as an algebraic semantics, period. They do not show
that this variety is an equivalent algebraic semantics for the
same logic — and of course they cannot do so, because there is
a unique quasivariety with this property, and such a slot is
already filled (as we have seen) by OML.
{\rm\cite[p.~189]{kowalski-paoli-giuntini-16}}}
\label{KPG6}
\end{claim}

\begin{response} {\it to Claim \ref{KPG6}}.
As a witness of their claims Kowalski, Paoli and Giuntini invoke the
{\em  equivalent algebraic semantics\/} (model) which favours OML
as the only acceptable model ({\em semantics}) of $\mathcal QL$.
That serves them to say that one cannot show that WOML is an
{\em equivalent algebraic semantics} (model) of $\mathcal QL$
because the ``slot is already filled by OML.'' Here, again, they
ignore the rigorously proved results of ours that $\mathcal QL$
has non-isomorphic and non-equivalent models and that OML is just
one of them, not occupying any privileged ``slot.'' 
\end{response}

\medskip 
We can sum up our responses to the claims made by
Kowalski, Paoli and Giuntini as follows:
\begin{itemize}
\item Kowalski, Paoli, and Giuntini's claim that one should use
  {\em  equivalent algebraic semantics\/} (model) which favours OML
  as the only acceptable model ({\em semantics}) of $\mathcal QL$
  is unfounded since we rigorously proved that all other
  models/semantics of $\mathcal QL$ stand on equal footing with OML
  in spite of the fact that they might be mutually non-equivalent
  and non-isomorphic---see Figure \ref{fig:two-tables} and Theorems
  \ref{th:compl-sound-womm}-\ref{th:compl-sound-Boole}
\item  Kowalski, Paoli, and Giuntini apparently object
  (although it is not quite clear from the statements they put
  forward in their paper) the fact that soundness and completeness
  of $\mathcal QL$ lead to the same set of theorems of $\mathcal QL$.
  Such an objection would be unfounded (if really held) because
  it is the essence of soundness and completeness of $\mathcal QL$
  with respect to all of its models/semantics in the sense that all
  its non-equivalent and non-isomorphic models must yield all their
  theorems and rules of inference, nothing more and nothing less. 
\item Kowalski, Paoli, and Giuntini claim that $\mathcal QL$
  is an {\em orthomodular\/} quantum logic is unfounded since a
  logic which is sound and complete with respect to non-orthomodular
  models (WOML*, WOML1*, \dots) can hardly be called orthomodular,
  especially since the orthomodularity equation given by
  Eq.~(\ref{om}) cannot be expressed within the $\mathcal QL$
  formalism. 
\end{itemize}

Another criticism comes from Drozdowska \cite{drozdowska-26}.
It exhibits misunderstandings and an error (which invalidates one
of her two main claims, \ref{D4}), but since responses to her
claims offer a good insight into the cardinal points of our results,
we reckon that the best way to address them would be in the
following Claim-and-Response manner. Drozdowska focuses on O6 and
matrix semantics, adequate matrices, mapping of O6 to a two-element
Boolean matrix, etc., but since O6 is a WOML, we make use of
the algebraic model language to address her criticism. 
Her claim that ``[the definitions of model and validity employed by
Pavi\v ci\'c and Megill are matrix-theoretic in character''
\cite[p.~304]{drozdowska-26} might only refer to O6 itself and/or
O6-valuation within the equivalence relation of Lindenbaum algebra
used to prove the completeness of $\mathcal CL$ and therefore in
our responses to her matrix-theoretic claims we keep to the abstract
algebraic language.

\begin{claim}{\rm (Drozdowska)} {\it\ 
There are many nontrivial unital matrices which are sound and
complete for Classical Logic, e.g. \dots weakly distributive
lattice O6. {\rm [But]} the existence of non-distributive matrix
models does not by itself provide evidence that classical logic
is non-distributive.
{\rm \cite[p.~284]{drozdowska-26}}.}
\label{D1}
\end{claim}

\begin{response} {\it to Claim \ref{D1}}.
Classical logic can hardly be called distributive since the
distributivity equation Eq.~(\ref{d}) cannot be expressed within
the $\mathcal CL$ formalism and since non-distributive models
exist.
\end{response}

\begin{claim}{\rm (Drozdowska)} {\it\  
One possible meaning of distributivity of $\mathcal CL$ concerns
the syntactic side of the logic. In this sense, a logic is
distributive if the distributivity laws
$A\wedge(B\vee C)\equiv_c(A\wedge B)\vee(A\wedge C)$
and $A\vee(B\wedge C)\equiv_c(A\vee B)\wedge(A\vee C)$ are its
theorems. {\rm \cite[p.~285]{drozdowska-26}}.}
\label{D2}
\end{claim}

\begin{response} {\it to Claim \ref{D2}}.
Such a definition of the distributivity law
(given by Eq.~(\ref{d-equiv}) does exist in the
literature \cite[p.~60]{quine}, but that traditionally assumes
that the only model of the classical logic is the Boolean algebra,
i.e., that the theorems are evaluated by means of $\{\top,\bot\}$
truth-tables \cite[p.~15]{quine}. This historical link of the
classical logic to distributivity should not be taken as a base
for calling classical logic {\em distributive\/} since by the
same token we can call it {\em orthomodular\/} because the
``orthomodularity law'' given by Eq.~(\ref{oml-equiv}) holds 
in it or {\em modular\/} because the ``modular law''
$\vdash_{\mathcal{CL}} (A\vee B)\wedge(A\vee C))
\equiv_q  (A\vee (B\wedge(A\vee C)))$ holds in it as well. On the
other hand, since models/semantics of $\mathcal CL$ can also be
non-distributive lattices, to call the logic {\em distributive\/}
is additionally not appropriate from a semantical point of view.
A possible approach to this terminological awkwardness would be
to restrain ourselves from attaching distributive, or orthomodular,
or modular appellation to classical logic altogether and simply call
$\mathcal CL$ a {\em classical logic\/}, instead. The denominations
distributivity, orthomodularity, and modularity which are ambiguous
in the classical logic should be reserved for its models, i.e.,
lattices where they have unequivocal equational meaning given by
Eq.~(\ref{d}), Eq.~(\ref{om}), and
$(a\cup b)\cap(a\cup c)=(a\cup(b\cap(a\cup c)))$), respectively.
(Notice that the weakly modular lattice is also a model of 
$\mathcal CL$.)
\end{response}

\begin{claim}{\rm (Drozdowska)} {\it \
{\rm [Firstly], [i]}n the case of O6, {\rm[Eq.~(\ref{d=1})]} does
not hold. Furthermore, formulas that are logically equivalent in
{\rm[$\mathcal CL$] [see Eq.~(\ref{d-equiv})]} are not semantically
equivalent {\rm[?]}$\!$ in matrix O6 {\rm[apparently referring to
violation of Eq.~(\ref{d=1}) in O6]}, and equivalence does not
have a normal {\rm[hic]} interpretation in matrix O6.}
\label{D3}
\end{claim}

\begin{response} {\it to Claim \ref{D3}}.
Here, again, as Kowalski, Paoli and Giuntini do, Drozdowska
finds it hard to accept the mathematical proof of the fact that
WOML, and therefore also O6 as its special instance, is a legitimate
model of $\mathcal CL$. In particular, she hints that the fact that 
existence of such a model is not {\em normal\/}, meaning that
Eq.~(\ref{d=1}) does not hold, might be a ``problem'' (in a
paragraph underneath this statement she writes: ``Secondly, another
``problem'' is that \dots'' \cite[p.~301]{drozdowska-26}). 
\end{response}

\begin{claim}{\rm (Drozdowska)} {\it\
Secondly, another problem is that in the lattice O6 the rule of
replacement of equivalents does not hold. The rule of replacement
of logical equivalents is a derivable rule of {\rm [$\mathcal CL$]}:
\begin{eqnarray}
  \frac{\Phi\equiv_c\Psi}{\Omega\equiv_c\Omega(\Phi/\Psi)},
  \label{eq:Omega}
\end{eqnarray}
which states that if a formula $\Phi$ is logically equivalent to
$\Psi$, then $\Omega$ is logically equivalent to a formula obtained
from $\Omega$ through replacement of one or more occurrences of
$\Phi$ in $\Omega$ by $\Psi$. If we take Pavi\v ci\'c and Megill’s
interpretation of the language of {\rm [$\mathcal CL$]} in
ortholattices, the rule will take the form:
\begin{eqnarray}
  \frac{\Phi\equiv_c\Psi=1}{\Omega\equiv_c\Omega(\Phi/\Psi)=1},
  \label{eq:Omega1}
\end{eqnarray}
Let $\Phi$ be the formula $A\vee(B\wedge C)$, $\Psi$ the formula
$((A\vee B)\wedge(A\vee C))$, and $\Omega$ the formula
$(A\vee(B\wedge C))\to_c(A\vee(B\wedge C))$. If we interpret these
formulas in O6 and replace the first occurrence of $\Phi$ (the
antecedent of $\Omega$) with $\Psi$, we obtain:
\begin{eqnarray}
\frac{(a\cup(b\cap c))\equiv_c((a\cup b)\cap(a\cup c))=1}
{((a\cup(b\cap c))\to_c(a\cup(b\cap c)))\equiv_c(((a\cup b)\cap(a\cup
c))\to_c(a\cup(b\cap c))) = 1}\nonumber
\end{eqnarray}
Consider the valuation assigning $a$ to $a$, $b$ to $b$, and $b'$
to $c$. Then the weak distributivity as the premise holds. In the
conclusion, however, the left side has the value:
$a\cup(b\cap b')\to_c(a\cup(b\cap b')=(a\cup 0)\to_c(a\cup 0)=
a\to_ca=a$, while the right side has value: $((a\cup b)\cap(a\cup
b'))\to_c(a\cup(b\cap b'))=(b\cap 1)\to_c(a\cup 0)=b\to_ca = 1$.
Hence, $a\neq 1$ {\rm[in the original text there is a typo
here: $a \equiv_c 1=a\neq 1$ was typed instead of $a\neq 1$]}.
Therefore, the rule fails in O6. This failure shows that O6 does
not preserve one of the fundamental structural properties of
classical logic, namely the substitutivity of logically equivalent
formulas. Consequently, although O6 may reproduce the set of valid
formulas, it does not fully preserve the inferential structure of CL.
\label{D4}}
\end{claim}

\begin{response} {\it to Claim \ref{D4}}.
This claim is wrong. It is based on an obvious error: $a\to_ca$ is
not equal to $a$ but to 1: $a\to_ca=a'\cup a=1$---see
Fig.~\ref{fig:Os}. Actually, we need not calculate the example to
arrive at the result at all. Every equation of the form
$f(a,b,\dots)=1$ passes O6. Only the equations of the form
$f(a,b,\dots)=g(a,b,\dots)$ can fail in O6. Consequently, the
claim ``that O6 does not fully preserve the inferential structure
of $\mathcal CL$'' has not been proved by Drozdowska. 
\end{response}

\medskip 
We can sum up our responses to the claims made by Drozdowska
as follows:
\begin{itemize}
\item Drozdowska's claim that the existence of non-distributive
  matrix models does not by itself provide evidence that
  classical logic $\mathcal CL$ is non-distributive and that it
  is distributive because the distributivity law (Eq.~(\ref{d}))
  is its theorem is unfounded because $\mathcal CL$ also has the
  orthomodularity law as well as the modular law as its theorems.
  Hence, classical logic $\mathcal CL$ could likewise be called
  an orthomodular or modular logic. In any case $\mathcal CL$ is
  neither non-distributive nor distributive---it is just
  {\em classical\/} logic.
\item Drozdowska's claim that the equivalence relation $\equiv_c$
  does not have an acceptable interpretation in O6 because formulas
  that are logically equivalent are not semantically equivalent
  is unfounded because all models map lattice formulas with a
  valuation ``1'' to provable theorems in the logic. That some
  models, like a distributive lattice DL, have equational formulas
  which other models do not have, does not affect this mapping.  
\item Drozdowska's claim that O6 does not fully preserve the
  inferential structure of $\mathcal CL$ is unfounded because it
  is based on an error in her calculation as explained in the
  Response of Claim  \ref{D4} and which boils down to the claimed
  failure of a formula which is equal to ``1'' in O6. However, all
  formulas that are equal to ``1'' in any of the models pass O6.
  Only those that are not, might fail in O6, like, e.g., the
  lattice distributivity given by Eq.~(\ref{d}).
\end{itemize}
  
\section{\label{Sec.~concl}Conclusion}

The goal of this paper and of revisiting of our previous results
in it is to show that the criticism of the results according to
which both quantum and classical logics have non-equivalent and
non-isomorphic models \cite[Chapters 8--11]{Schechter},
\cite{mpcommp99,pm-ql-l-hql1,pavicic-amp-16}, attempted by
Kowalski, Paoli and Giuntini \cite{kowalski-paoli-giuntini-16}
and Drozdowska \cite{drozdowska-26}, are unfounded and/or erroneous.

In order to reach the goal, we start with a renewed side-by-side
presentation of logics in Sec.~\ref{Sec.~QandCL} and non-equivalent
and non-isomorphic lattices in Sec.~\ref{Sec.~QandCLa}. Then, in
Sec.~\ref{Sec.~models} we show that those non-equivalent and
non-isomorphic lattices are models of the logics and prove
soundness and completeness for all of them which we summarise
in Fig.~\ref{fig:two-tables}, again in a side-by-side fashion.
Notice that Secs.~\ref{Sec.~QandCL}-\ref{Sec.~models} material
are from Refs.~\cite{mpcommp99,pm-ql-l-hql1,pavicic-amp-16}, just
presented in a renewed manner so as to make the main result that
the same logics might have non-equivalent and non-isomorphic models
more accessible and digestible to the reader and less likely to
be misunderstood in the future, again. 

We consider, analyse, and respond to the criticisms in
Sec.~\ref{Sec.~crit} made by Kowalski, Paoli and Giuntini
\cite{kowalski-paoli-giuntini-16} and by Drozdowska
\cite{drozdowska-26}. The former one deals mostly with
quantum logic and the latter mostly with classical logic. 

Kowalski, Paoli and Giuntini try to legitimate the orthomodular
lattice as the only semantics/model of quantum logic with the
following argumentation: ``whether or not [quantum logic] adequately
represents a proper formalisation of the intrinsic logic of quantum
mechanics, its algebraic counterpart can’t be anything else than the
variety of orthomodular lattices.''
\cite[p.~188]{kowalski-paoli-giuntini-16} 
Hence, they overlook the fact that all models be they mutually
non-equivalent and non-isomorphic or not induce the same sets of
theorems, axioms, and rules of inferences which represent quantum
logic and that therefore the {\em can’t-be-anything-else} argument
is unfounded. The fact that only one of the models underlies the
structure of quantum mechanics has nothing to do with the existence
of other models. On the other hand, their claim that quantum logic
itself is orthomodular is unfounded since the lattice
orthomodularity equation cannot be expressed within the logical
formalism.

Drozdowska's criticism boils down to three claims. The first one
is an attempt, similar to the one of Kowalski, Paoli and Giuntini,
to establish the distributivity trait via the non-equational
distributivity law. However, this claim is unfounded since the
classical logic also has the non-equational forms of the
orthomodularity law as well as the modular law as its theorems.

The second claim that it is a ``problem'' that the formulas that
are logically equivalent are not semantically equivalent is
unfounded because all models map lattice formulas with a valuation
“1” to provable theorems in the logic. The fact that the
distributive lattice (Boolean algebra) has equational formulas
which other models do not have, does not affect the ``1'' mapping
of formulas of non-equivalent models to the logic. They all map to
to the same set of logic theorems.

Third Drozdowska's claim, i.e., her ``proof'' that O6 does not
fully preserve the inferential structure of classical logic is
simply wrong. The claim stemmed from an error she made in a
calculation as explained in the Response to Claim \ref{D4} in
Sec.~\ref{Sec.~crit}. Actually, as it refers to a violation of a
lattice formula equal to ``1'' in O6, Drozdowska should not have
calculated anything at all since all such equations pass O6.

Both criticisms give an impression of being written in an
indecisive manner as if the authors just were not sure how to
proceed with moulding their dissatisfaction with and analogous
results from the literature. So, we are drawn to the conclusion
that it might be more productive to commit ourselves to solving
open problems in the field of orthomodular lattices, instead.
Fifteen of them are, for instance, given in
\cite[p.~789,90]{pm-ql-l-hql2}.


\end{document}